\documentclass[aps,twocolumn,prl,tightenlines,floatfix,showpacs]{revtex4-1}
\usepackage{graphicx}
\usepackage{amsmath}
\usepackage{amssymb}
\usepackage{times}
\usepackage[english]{babel}

\begin{document}

\title{Bosonic and fermionic transport phenomena of ultra-cold atoms in 1D optical lattices}
\author{Chih-Chun Chien$^{1}$, Michael Zwolak$^{2}$, and Massimiliano Di Ventra$^{3}$}

\affiliation{$^{1}$Theoretical Division, Los Alamos National Laboratory, MS B213, Los Alamos, NM 87545, USA \\
$^{2}$Department of Physics, Oregon State University, Corvallis, OR 97331, USA \\
$^{3}$Department of Physics, University of California, San Diego, CA 92093, USA}

\date{\today}

\begin{abstract}
Using the micro-canonical picture of transport -- a framework ideally suited to describe the dynamics of closed quantum systems such as ultra-cold atom experiments -- we show that the exact dynamics of non-interacting fermions and bosons exhibit very different transport properties when the system is set out of equilibrium by removing the particles from half of the lattice. We find that fermions rapidly develop a finite quasi steady-state current reminiscent of electronic transport in nanoscale systems. This result is robust -- it occurs with or without a
harmonic confining potential and at zero or finite temperature. The zero-temperature bosonic current instead exhibits strong oscillatory behavior that decays into a  steady-state of zero current only in the thermodynamic limit. These differences appear most strikingly in the different particle number fluctuations on half of the lattice as a consequence of the spin statistics. These predictions can be readily verified experimentally.
\end{abstract}


\maketitle

Ultra-cold atoms loaded in various optical lattices help visualize and gain a deeper understanding of many exciting quantum phenomena. For instance, Hanbury-Brown-Twiss types of experiments measuring spatial correlations of density fluctuations in non-interacting bosons \cite{boson_bunching} and fermions \cite{fermion_antibunching} clearly demonstrate bunching and anti-bunching effects, which are definite signatures of the underlying spin statistics. On the other hand, bosonic and fermionic Mott insulators \cite{bosonMott,fermionMott} have been realized experimentally and the statistics do not seem to play an important role in their thermodynamic properties if magnetic phenomena due to multi-component fermions are not considered. It is then clear that the question of how and when statistics affect an atomic system is of great fundamental interest and worth further investigations.

In addition, recent progress in experimental techniques of ultra-cold atoms allows us to probe not only static structures, but also the dynamics of atoms in optical lattices. There have been experiments \cite{collision_tran,interaction_tran} showing that interactions (via collisions of atoms) can affect the motion of an atomic cloud when released from a non-equilibrium position. It has also been shown that mass and entropy transport of bosons in optical lattices can be unexpectedly slow \cite{slowmasstran}. Adding to the excitement are the experiments of Ref.~\cite{Greiner_spin} which show how spin dynamics and phase transitions can be simulated using ultra-cold atoms. The possibility of using these systems to simulate electronic devices and developing atom-based devices has also opened another research direction named ``atomtronics'' \cite{atomtronics}. The often parallel investigations of spin-statistics and transport have therefore gained momentum in the study of ultra-cold atoms in optical lattices.

Here we explore the effect of the underlying statistics on atomic transport phenomena. Since the deviation from the equilibrium could be significant, the applicability of the Kubo formula may be limited. Moreover, the atomic cloud in a real experiment should be modeled practically as a closed system. This could invalidate the popular Landauer formalism \cite{Landauer} which connects the system to reservoirs to study its static transport properties. Instead, we base our theoretical studies on the micro-canonical formalism (MCF) of transport introduced in Ref.~\cite{Micro} to study particle dynamics in nanoscale systems (see also Refs.~\cite{MCE,Maxbook}). Specifically, this formalism explores transport properties of a closed quantum system where the dynamics of a finite number of particles -- whether interacting or not -- are followed in time~\cite{Maxbook}. MCF is thus particularly suited for simulating transport in ultra-cold atomic systems as done in actual experiments. This formalism allows us to make specific predictions on the different roles fermion and boson statistics play during the dynamical evolution of atomic gases. 

\begin{figure}
  \includegraphics[width=1.8in,clip]{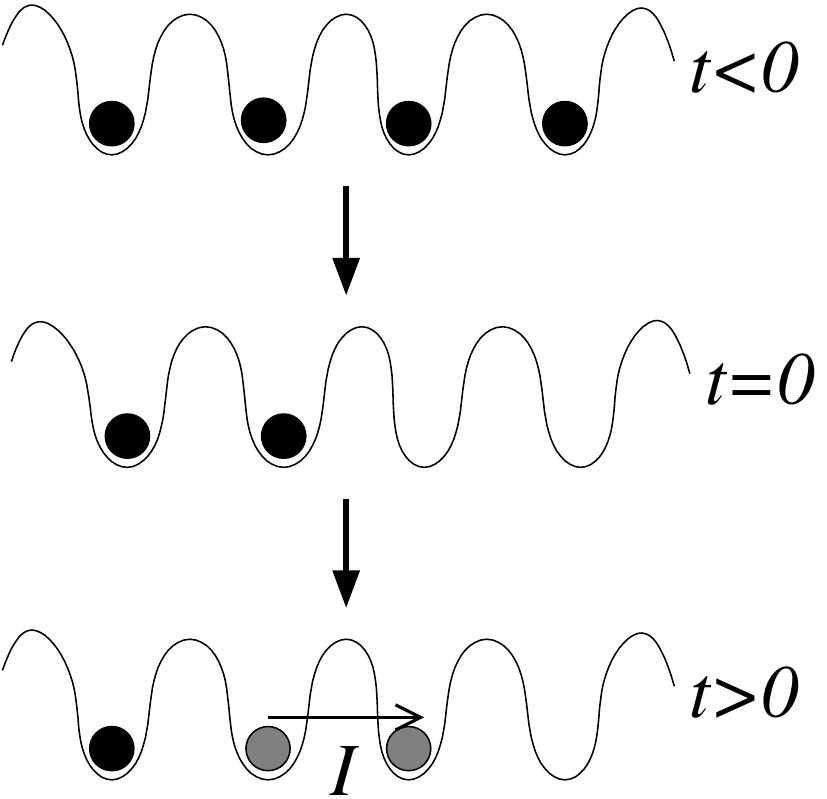}
  \caption{Schematic plot of the set up for fermions that we simulate. The gray dots emphasize that atoms may be in a superposition of locations.}
\label{fig:cartoon}
\end{figure}

Here we consider the following experimental situation that may be realized in 1D optical lattices -- even though generalizations to higher dimensions are possible \cite{Micro}. We consider non-interacting particles whose initial state corresponds to a given filling $n$.
For a one-band model with moderate lattice depth, if the atoms are non-interacting fermions (bosons) of a single species, the ground state is a band insulator (quasi-Bose-Einstein condensate with all atoms in the lowest energy state). Then we remove part of the atoms -- say, those on the right half of the lattice. This can be realized by a sudden excitation with a focused laser beam so that those atoms leave the trap while the lattice remains the same. This creates a vacuum region which then induces a current. Figure~\ref{fig:cartoon} demonstrates this experimental setup.

In the MCF, the system is initially prepared in the ground state of one Hamiltonian $H_{0}$ and then it evolves according to a second Hamiltonian $H_{e}$ of which it is not an eigenstate~\cite{Micro}.
We formulate the evolution in terms of the correlation matrix $c_{ij}(t)=\langle GS_0|c^{\dagger}_{i}(t)c_{j}(t)|GS_0\rangle$, where $c^{\dagger}_{i}$ ($c_{i}$) denotes the creation (annihilation) operator of a particle at site $i$ and  $|GS_0\rangle$ denotes the ground state of $H_0$ (compare to Ref.~\cite{MCE} where the single particle states were used). For a system of $N$ sites, the current through the middle is $
I=- d\langle \hat{N}_L(t)\rangle/dt$.
Here $\hat{N}_L(t)=\sum_{i=1}^{N/2}c^{\dagger}_{i}(t)c_{i}(t)$ counts the number of particles on the left half of the lattice. To measure the current, one may follow Ref.~\cite{slowmasstran} and prepare several identical systems, then takes images at different times. The current is the rate at which atoms are transferred from the left to the right.

To determine the initial state, we consider a tight-binding hopping Hamiltonian of non-interacting particles, $H_{0}=-\tilde{t}\sum_{\langle ij\rangle}c^{\dagger}_{i}c_{j}+\sum_{j}(1/2)m\omega^{2}d^{2}(j-N/2-1/2)^{2}c_{j}^{\dagger}c_{j}$,
on a finite, non-periodic lattice. Here, $\tilde{t}$ is the hopping coefficient and it determines the unit of time, $t_0\equiv \hbar/\tilde{t}$ and $m$, $\omega$, $d$ are the mass of atoms, the trap frequency of a background harmonic potential, and the lattice constant, respectively. We define $Y\equiv m\omega^{2}d^{2}/\tilde{t}$ with $Y\sim 0.001$ as in Refs.~\cite{collision_tran,interaction_tran}. For fermions of one species, s-wave scattering and the Pauli exclusion principle naturally forbid observable interactions among atoms. For bosons we assume the background scattering is negligible. We remark that no dissipation mechanism is considered in our study, which is consistent with Refs.~\cite{collision_tran,interaction_tran}.

The system is filled up to the given filling factor $n$ and the ground state is best described in the energy basis with the creation (annihilation) operator $b^{\dagger}_{k}$ ($b_{k}$). Here the index $k$ labels different energy eigenvalues and should not be confused with the momentum. The unitary transformation $c_{j}=\sum_{k}(U)_{jk}b_{k}$ diagonalizes the Hamiltonian as $H_{h}=\sum_{k=1}^{N}\epsilon_{k}b^{\dagger}_{k}b_{k}$. When $Y=0$, one has $\epsilon_k = -2 \tilde{t} \cos[k\pi/(N+1)]$ and $U_{jk}=\sqrt{\frac{2}{N+1}}\sin(\frac{jk\pi}{N+1})$ with $j,k=1,\cdots, N$. For $t<0$ the correlation function at zero temperature ($T=0$) is $\langle b^{\dagger}_{k}b_{k^{\prime}}\rangle=\delta_{kk^{\prime}}$ for fermions and $\langle b^{\dagger}_{k}b_{k^{\prime}}\rangle=N\delta_{kk^{\prime}}\delta_{k1}$ for bosons, where $k=1$ denotes the lowest energy state. For fermions of one species, at $n=1$ the lowest band is completely filled, thus there are no spatial correlations for fermions due to Pauli exclusion principle, but there are ample spatial correlations for bosons. At finite $T$, $\langle b^{\dagger}_{k}b_{k^{\prime}}\rangle=n_{k}\delta_{kk^{\prime}}$, where $n_{k}=(\exp[(\epsilon_{k}-\mu)/k_{B}T]\pm 1)^{-1}$ is the corresponding thermal distribution and $\mu$ is the chemical potential. We set $k_{B}\equiv 1$.

\begin{figure}
  \includegraphics[width=3.in,clip]{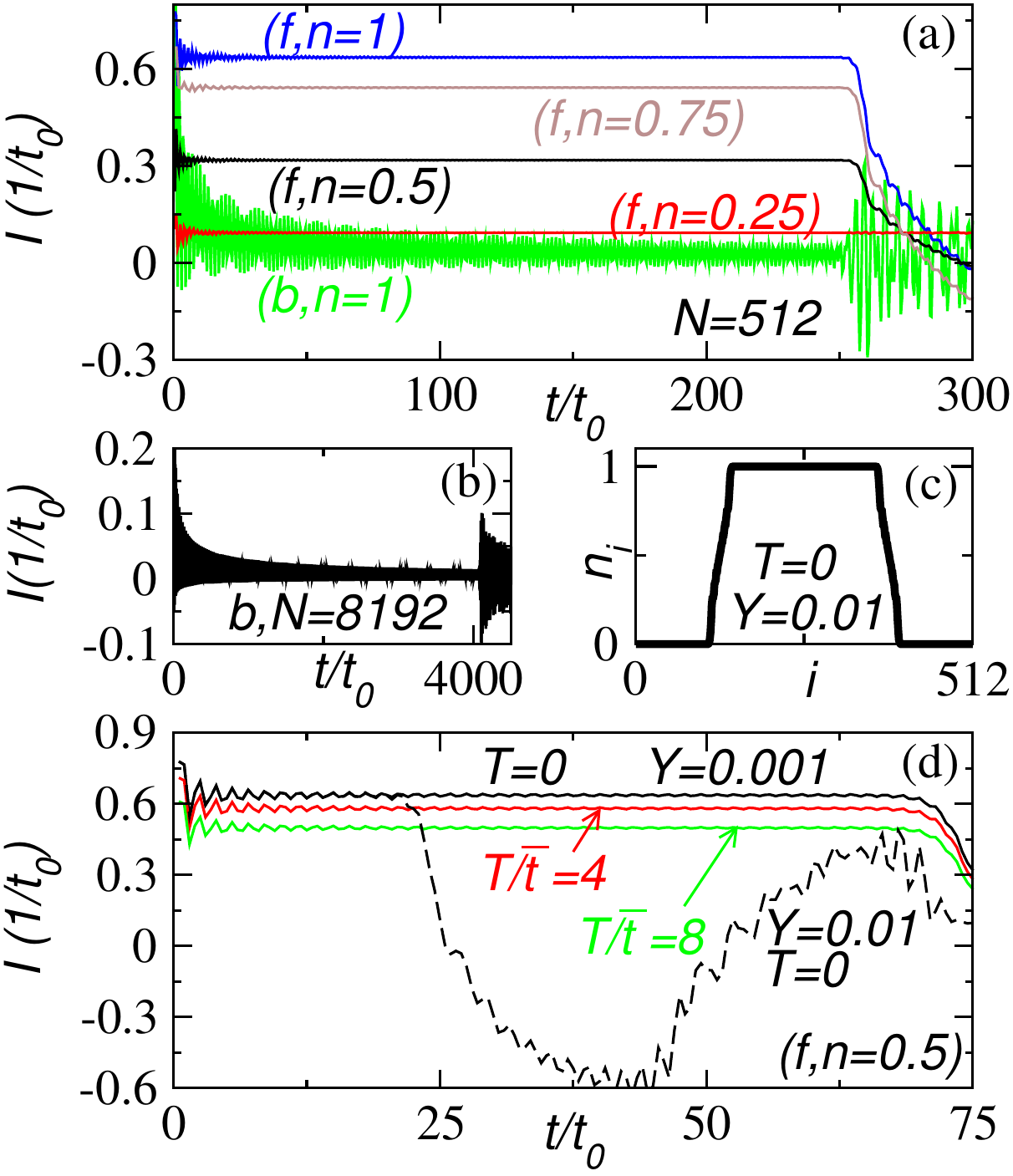}
  \caption{(Color online) (a) Currents in a uniform lattice for fermions (top) and bosons (bottom) with initial $n=1$ at $T=0$ after the atoms on the right are removed. The fermionic currents for initial fillings $n=0.75, 0.5, 0.25$ are also shown (labelled next to each curve). Here $N=512$. (b) The bosonic current for $N=8192$ with initial filling $n=1$. (c) The density profile for fermions with $N=512$, $n=0.5$, and $Y=0.001$ at $T=0$. (d) Fermionic currents in a lattice with a harmonic potential for $n=0.5$ and $N=512$. The solid lines correspond to $T/\tilde{t}=0, 4, 8$ with $Y=0.001$ while the dashed line corresponds to $T/\tilde{t}=0$ with $Y=0.01$. Here $Y=m\omega^{2}d^{2}/\tilde{t}$. }
\label{fig:n1}
\end{figure}

The correlation matrix in real space can be written down using the unitary transformation as $c_{ij}=\sum_{k,k^{\prime}}(U^{\dagger})_{ki}\langle b^{\dagger}_{k}b_{k^{\prime}}\rangle U_{jk^{\prime}}$. In Fig.~\ref{fig:n1}(c) we show the density profile of the ground state with $n=0.5$ initially and $Y=0.001$ at $T=0$, which resembles the wedding-cake structure of bosonic Mott insulators \cite{slowmasstran}. At $t=0$ all atoms on the right half are blown off by a laser beam whose frequency is different from that of the laser generating the optical lattices so that the lattice remains intact. The consequence is that $c_{ij}(t=0)$ becomes zero if $i$ or $j$ is on the right half, but the submatrix of the left half (both $i,j\le N/2$) is not affected.

The evolution of $c_{ij}$ is determined by
\begin{eqnarray}\label{eq:cij}
c_{ij}(t)&=&\sum_{k,k^\prime=1}^{N}(U^{\dagger})_{ki}(U)_{jk\prime}D_{kk^\prime}(0)e^{i(\epsilon_{k}-\epsilon_{k^{\prime}})t}.
\end{eqnarray}
Here $D_{kk^\prime}(0)=\sum_{i,j=1}^{N/2}(U^{\dagger})_{k^\prime j^\prime}U_{ik}c_{ij}(t=0)$ is the initial correlation matrix in energy basis after the atoms on the right half are removed. The current from the left to the right is then
\begin{equation}\label{eq:Icij}
I=-2\tilde{t}\mbox{Im}\langle c^{\dagger}_{N/2}(t)c_{N/2+1}(t)\rangle.
\end{equation}
Although bosons and fermions obey different statistics, the expression for the respective current is the same. The behavior of the current, however, is qualitatively different.

Figure~\ref{fig:n1}(a) shows the currents of fermions and bosons on a uniform lattice of $N=512$ with initial $n=1$ at $T=0$ after the atoms on the right half are removed.
The fermionic current reaches a plateau after a short initial transient time. This constant current has a duration proportional to the system size, indicating that fermions will develop a true steady state of finite current in the thermodynamic limit. For the finite-system case, the current varies abruptly when the electrons traveling to the right reach the far boundary and come back, as is seen in Fig.~\ref{fig:n1}(a). Here we will focus on the physics prior to this finite-size revival behavior. The almost constant current found here is similar to the quasi steady-state current (QSSC) shown in Ref.~\cite{MCE} where a step-function bias is applied to a half-filled lattice. As in that case, it shows that a quasi steady state can originate dynamically in the {\it absense} of interactions, solely due to the wave properties of the
system~\cite{Micro}. Moreover, QSSCs can develop in higher dimensional geometries, as shown in Ref.~\cite{MCE}.

Importantly, the fermionic QSSC is observable in the presence of a reasonable harmonic trap potential for both $T=0$ and finite $T$ \cite{munote} as shown in Fig.~\ref{fig:n1}(d). We note that the length of the initial transient time during which the current oscillates significantly depends on the initial state configuration and how the system is set out of equilibrium.

In contrast, the $T=0$ bosonic current exhibits rapid oscillations with a period $\approx t_0$ ($\Delta I/I_{avg}\gg 1$, where $\Delta I$ is the amplitude of the oscillations and $I_{avg}$ is the current averaged over a period of time). Therefore, a QSSC is not found \cite{bosonTnote}. However, as it is also
evident from Fig.~\ref{fig:n1}(a) and (b) that the oscillations of the bosonic current decrease with time.
For bosons with initial filling $n=1$ on a uniform lattice of size $N$, the correlation matrix in energy basis after the atoms on the right half lattice are removed is $D_{kk^{\prime}}(t=0)=N\sum_{i,j\le N/2}U_{i1}U_{ji}U_{ik}U_{jk^{\prime}}$ at $T=0$. In the limit $N\rightarrow\infty$ it approaches $NF(p)F(p^{\prime})$,
where $F(p)=p\cos(p\pi/2)/[(1-p^{2})\pi]$. $F(p=\textrm{odd})=0$, except $F(1)=1/4$ and $F(p=\textrm{even})$ is finite.

The current from Eq.~\eqref{eq:Icij} then becomes
\begin{eqnarray}
I&=&2\tilde{t}\sum_{p\neq p^{\prime}}\sin\left\{2\tilde{t}t\left[\cos(\frac{p\pi}{N+1})-\cos(\frac{p^{\prime}\pi}{N+1})\right] \right\}\times \nonumber \\
& &U_{N/2,p}U_{N/2+1,p^{\prime}}NF(p)F(p^{\prime}).
\end{eqnarray}
In the final expression only terms with $\{p=1, p^{\prime}=\textrm{even}\}$ and $\{p^{\prime}=1,p=\textrm{even}\}$ contribute and their contributions are equal. We focus on the long-time limit and set $t=(N/2)t_0$. One can make other choices as long as $t/t_{0}\sim O(N/2)$ and obtain the same conclusions. We consider $\{p=1, p^{\prime}=\textrm{even}\}$ and the final result is simply twice the current of this part. Due to the rapid oscillatory part of the first term for large $t/t_0$, only $0<p^{\prime}/(N+1)<1/\sqrt{N}$ may contribute finitely. The current then approaches
\begin{eqnarray}
I&\rightarrow&\frac{-\tilde{t}}{2\pi}\int_{0}^{\frac{1}{\sqrt{N}}}d\bar{p}\sin[N(1-\cos(\bar{p}\pi)]\sin\left(\frac{\bar{p}\pi}{2}\right)\frac{1}{\bar{p}} \propto N^{-1/2} \nonumber
\end{eqnarray}
Therefore in the thermodynamic limit, the long-time current decays to zero for bosons. This analysis is supported by our simulations shown in Fig.~\ref{fig:n1}(a) and (b).


Another interesting question is how the filling factor affects the QSSC for fermions. If one performs the experiments separately for different initial filling factors $n=1$, $0.75$, $0.5$, and $0.25$ on a uniform lattice using fermions, the QSSC does not exhibit  linear dependence on $n$ as shown in Fig.~\ref{fig:n1}(a). This non-linear dependence of the QSSC versus the filling factor originates from the band structure, which results in differing amounts of correlations between sites on the lattice. As the filling gets lower, the initially remaining fermions on the left half lattice are more correlated because the constraint from Pauli exclusion principle is less restrictive. This correlation effect competes with the counting of contributions from different energy states.
However, the initial filling does not affect the qualitative behavior of the QSSC of fermions if it is not too different from $n=1$.

Before discussing other differences of the two dynamics, let us first check if they can be observed experimentally. For optical lattices with moderate depth, with $V_0$ and $E_R$ the lattice depth and recoil energy, respectively, a typical magnitude of the hopping parameter $\tilde{t}$ can be estimated from the parameters given in Ref.~\cite{Holattice}. For $V_0/E_R=10$, $\tilde{t}\approx 3$ nK which corresponds to $t_{0}=2.5$ ms. Therefore the oscillations of the bosonic current are not too difficult to detect. Moreover, the typical holding time for recent experiments can be of the order of one second \cite{collision_tran} so the QSSC of fermions, which is of the order of about $50t_{0}\sim 0.13$s for a lattice of $N=512$ sites with a reasonable harmonic potential, should also be observable for $T\le 8\tilde{t}$ according to Fig.~\ref{fig:n1}(d).

We now show that particle number fluctuations can distinguish the underlying spin statistics when the system is {\it dynamically evolving}. Let us evaluate the instantaneous fluctuations in the total number of particles on the left half lattice, $\Delta N^{2}$. Experimentally $\Delta N^{2}$ may be measured by repeating the preparation and measurement at a given time so the fluctuations could be inferred. By definition,
\begin{equation}
\Delta N^2=\langle \hat{N}_{L}^{2} \rangle-\langle \hat{N}_{L} \rangle^{2}.
\end{equation}
Here $\langle \hat{N}_{L}^{2} \rangle=\sum_{i=1}^{N/2}\langle\hat{n}_{i}^{2}\rangle+2\sum_{i<j}^{N/2}\langle\hat{n}_{i}\hat{n}_{j}\rangle$, where $\hat{n}_{i}=c^{\dagger}_{i}c_{i}$ and $n_{i}=\langle \hat{n}_{i}\rangle$. This expression involves expectation values of four operators and one can decompose them using Wick's theorem. The decomposition, however, is sensitive to the underlying spin statistics. After some algebra, one has
\begin{eqnarray}
\Delta N^2_{\mbox{fermions}}&=&\sum_{i=1}^{N/2}n_{i}(1-n_{i})-2\sum_{i<j}^{N/2}|c_{ij}|^{2} ,  \label{eq:c2f} \\
\Delta N^2_{\mbox{bosons}}&=&\sum_{i=1}^{N/2}n_{i}(1+n_{i})+2\sum_{i<j}^{N/2}|c_{ij}|^{2}. \label{eq;c2b}
\end{eqnarray}
In contrast to the current, the expressions for the fluctuations are different for fermions and bosons. In the following we focus on $T=0$.

For fermions, $0\le n_i, |c_{ij}|\le 1$ due to Pauli exclusion principle. Clearly, the correlation terms $|c_{ij}|^{2}$ \textit{reduce} the number fluctuations. For the initial filling $n=1$, right after the particles on the right half lattice are removed the instantaneous number fluctuations on the left half are zero because Pauli exclusion principle forces one particle per site. In contrast, the initial boson distribution does not have this constraint. Thus by removing particles on the right half lattice, one removes half of the total number of particles \textit{on average} so initial number fluctuations are present. Moreover, $|c_{ij}|^2$ contributes positively to the fluctuations of bosons.

\begin{figure}
 \includegraphics[width=3.4in,clip]{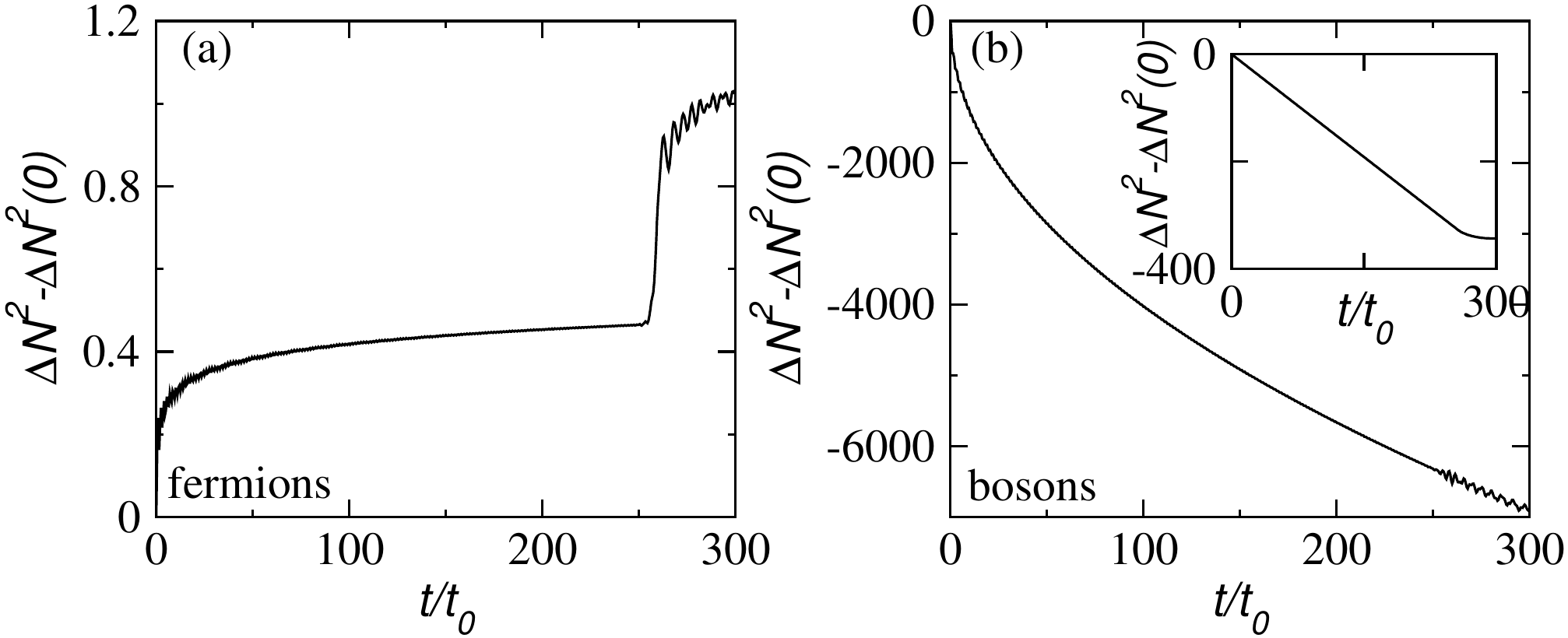}
  \caption{Particle number fluctuations on the left half lattice for (a) fermions and (b) bosons in a uniform lattice. Here $n=1$ before the particles on the right half are removed and $N=512$. The inset shows the bosonic fluctuations of a hypothetical case where the initial configuration is identical to that of the fermions.}
\label{fig:c2}
\end{figure}
To better visualize these results we plot in Fig.~\ref{fig:c2} the quantity $\Delta N^{2}(t)-\Delta N^{2}(t=0)$ for fermions and bosons in a uniform lattice. For fermions the fluctuations increase slowly after an initial transient time. This slow change in fluctuations corresponds to the QSSC in Fig.~\ref{fig:n1} (a). The behavior for bosons is very different. The contribution from spatial correlation, $|c_{ij}|^{2}$, decreases with time so the fluctuations decrease  accordingly. The initial state of bosons is a quasi condensate so there is strong coherence among particles on different sites. As the system evolves and more particles move to the right, the coherence decreases rapidly so the number fluctuations are reduced when compared to the initial number fluctuations. Since there are about $N^2$ terms in the contribution from spatial correlations, the observed change in $\Delta N^{2}$ for bosons is several orders
of magnitude larger than that for the fermions.

There are important differences when comparing our results with the Hanbury-Brown-Twiss type of experiments \cite{boson_bunching, fermion_antibunching}, where spatial correlations of density fluctuations are measured. While those  experiments \cite{boson_bunching, fermion_antibunching} utilize the underlying spin statistics for an \textit{equilibrium} system but implement a dynamical measurement (time-of-flight imaging), our discussion depends on the statistics of the initial distribution as well as \textit{the corresponding quantum dynamics}. Thus the difference in $\Delta N^{2}$ demonstrated here is a \textit{dynamic} property.

To further stress this point, we consider a hypothetical case for bosons where the initial configuration is assumed to be exactly the same as that of fermions, i.e., the lowest $N$ energy states for $n=1$ are occupied by one boson each before the particles on the right half are removed. Due to this selected initial correlation matrix, the bosonic current will be exactly the same as the fermionic current shown in Fig.~\ref{fig:n1}(a) and will show a QSSC. However, the bosonic number fluctuations, as shown in the inset of Fig.~\ref{fig:c2}(b), are very different from the fermionic ones shown on Fig.~\ref{fig:c2}(a) so the underlying statistics can still be resolved.

We conclude that non-equilibrium dynamics of non-interacting atomic gases already demonstrates unexpected phenomena. Ultra-cold atoms in optical lattices have the great potential of demonstrating a QSSC and its dependence on the spin statistics in non-interacting clean systems, which is difficult to realize for electronic systems due to Coulomb interactions. Since interactions between atoms are experimentally tunable, we anticipate that transport phenomena in the presence of interactions will also reveal interesting physics \cite{TormaPRL11}. The setup discussed here could also be of interest to the community of quench dynamics \cite{PolkovnikovRMP}. Therefore the micro-canonical picture of transport presented here could contribute significantly to our understanding of many fundamental aspects of transport phenomena in ultra-cold atoms.

We thank Prof. B. DeMarco for useful discussions. CCC acknowledges the support of the U. S. DOE through the LANL/LDRD Program.
MD acknowledges support from the DOE grant DE-FG02-05ER46204 and UC Laboratories.

\bibliographystyle{apsrev4-1}

\begin{thebibliography}{18}%
\makeatletter
\providecommand \@ifxundefined [1]{%
 \@ifx{#1\undefined}
}%
\providecommand \@ifnum [1]{%
 \ifnum #1\expandafter \@firstoftwo
 \else \expandafter \@secondoftwo
 \fi
}%
\providecommand \@ifx [1]{%
 \ifx #1\expandafter \@firstoftwo
 \else \expandafter \@secondoftwo
 \fi
}%
\providecommand \natexlab [1]{#1}%
\providecommand \enquote  [1]{``#1''}%
\providecommand \bibnamefont  [1]{#1}%
\providecommand \bibfnamefont [1]{#1}%
\providecommand \citenamefont [1]{#1}%
\providecommand \href@noop [0]{\@secondoftwo}%
\providecommand \href [0]{\begingroup \@sanitize@url \@href}%
\providecommand \@href[1]{\@@startlink{#1}\@@href}%
\providecommand \@@href[1]{\endgroup#1\@@endlink}%
\providecommand \@sanitize@url [0]{\catcode `\\12\catcode `\$12\catcode
  `\&12\catcode `\#12\catcode `\^12\catcode `\_12\catcode `\%12\relax}%
\providecommand \@@startlink[1]{}%
\providecommand \@@endlink[0]{}%
\providecommand \url  [0]{\begingroup\@sanitize@url \@url }%
\providecommand \@url [1]{\endgroup\@href {#1}{\urlprefix }}%
\providecommand \urlprefix  [0]{URL }%
\providecommand \Eprint [0]{\href }%
\@ifxundefined \urlstyle {%
  \providecommand \doi  [0]{\begingroup \@sanitize@url \@doi}%
  \providecommand \@doi [1]{\endgroup \@@startlink {\doibase
  #1}doi:\discretionary {}{}{}#1\@@endlink }%
}{%
  \providecommand \doi  [0]{doi:\discretionary{}{}{}\begingroup
  \urlstyle{rm}\Url }%
}%
\providecommand \doibase [0]{http://dx.doi.org/}%
\providecommand \Doi [0]{\begingroup \@sanitize@url \@Doi }%
\providecommand \@Doi  [1]{\endgroup\@@startlink{\doibase#1}\@@Doi}%
\providecommand \@@Doi [1]{#1\@@endlink}%
\providecommand \selectlanguage [0]{\@gobble}%
\providecommand \bibinfo  [0]{\@secondoftwo}%
\providecommand \bibfield  [0]{\@secondoftwo}%
\providecommand \translation [1]{[#1]}%
\providecommand \BibitemOpen [0]{}%
\providecommand \bibitemStop [0]{}%
\providecommand \bibitemNoStop [0]{.\EOS\space}%
\providecommand \EOS [0]{\spacefactor3000\relax}%
\providecommand \BibitemShut  [1]{\csname bibitem#1\endcsname}%
\bibitem [{\citenamefont {Folling}\ \emph {et~al.}(2005)\citenamefont
  {Folling}, \citenamefont {Gerbier}, \citenamefont {Widera}, \citenamefont
  {Mandel}, \citenamefont {Gericke},\ and\ \citenamefont
  {Bloch}}]{boson_bunching}%
  \BibitemOpen
  \bibfield  {author} {\bibinfo {author} {\bibfnamefont {S.}~\bibnamefont
  {Folling}}, \bibinfo {author} {\bibfnamefont {F.}~\bibnamefont {Gerbier}},
  \bibinfo {author} {\bibfnamefont {A.}~\bibnamefont {Widera}}, \bibinfo
  {author} {\bibfnamefont {O.}~\bibnamefont {Mandel}}, \bibinfo {author}
  {\bibfnamefont {T.}~\bibnamefont {Gericke}}, \ and\ \bibinfo {author}
  {\bibfnamefont {I.}~\bibnamefont {Bloch}},\ }\href@noop {} {\bibfield
  {journal} {\bibinfo  {journal} {Nature},\ }\textbf {\bibinfo {volume}
  {434}},\ \bibinfo {pages} {481} (\bibinfo {year} {2005})}\BibitemShut
  {NoStop}%
\bibitem [{\citenamefont {Rom}\ \emph {et~al.}(2006)\citenamefont {Rom},
  \citenamefont {Best}, \citenamefont {van Oosten}, \citenamefont {Schneider},
  \citenamefont {Folling}, \citenamefont {Paredes},\ and\ \citenamefont
  {Bloch}}]{fermion_antibunching}%
  \BibitemOpen
  \bibfield  {author} {\bibinfo {author} {\bibfnamefont {T.}~\bibnamefont
  {Rom}}, \bibinfo {author} {\bibfnamefont {T.}~\bibnamefont {Best}}, \bibinfo
  {author} {\bibfnamefont {D.}~\bibnamefont {van Oosten}}, \bibinfo {author}
  {\bibfnamefont {U.}~\bibnamefont {Schneider}}, \bibinfo {author}
  {\bibfnamefont {S.}~\bibnamefont {Folling}}, \bibinfo {author} {\bibfnamefont
  {B.}~\bibnamefont {Paredes}}, \ and\ \bibinfo {author} {\bibfnamefont
  {I.}~\bibnamefont {Bloch}},\ }\href@noop {} {\bibfield  {journal} {\bibinfo
  {journal} {Nature},\ }\textbf {\bibinfo {volume} {444}},\ \bibinfo {pages}
  {733} (\bibinfo {year} {2006})}\BibitemShut {NoStop}%
\bibitem [{\citenamefont {Greiner}\ \emph {et~al.}(2002)\citenamefont
  {Greiner}, \citenamefont {Mandel}, \citenamefont {Esslinger}, \citenamefont
  {Hansch},\ and\ \citenamefont {Bloch}}]{bosonMott}%
  \BibitemOpen
  \bibfield  {author} {\bibinfo {author} {\bibfnamefont {M.}~\bibnamefont
  {Greiner}}, \bibinfo {author} {\bibfnamefont {O.}~\bibnamefont {Mandel}},
  \bibinfo {author} {\bibfnamefont {T.}~\bibnamefont {Esslinger}}, \bibinfo
  {author} {\bibfnamefont {T.~W.}\ \bibnamefont {Hansch}}, \ and\ \bibinfo
  {author} {\bibfnamefont {I.}~\bibnamefont {Bloch}},\ }\href@noop {}
  {\bibfield  {journal} {\bibinfo  {journal} {Nature},\ }\textbf {\bibinfo
  {volume} {415}},\ \bibinfo {pages} {39} (\bibinfo {year} {2002})}\BibitemShut
  {NoStop}%
\bibitem [{\citenamefont {Jordens}\ \emph {et~al.}(2008)\citenamefont
  {Jordens}, \citenamefont {Strohmaier}, \citenamefont {Gunter}, \citenamefont
  {Moritz},\ and\ \citenamefont {Esslinger}}]{fermionMott}%
  \BibitemOpen
  \bibfield  {author} {\bibinfo {author} {\bibfnamefont {R.}~\bibnamefont
  {Jordens}}, \bibinfo {author} {\bibfnamefont {N.}~\bibnamefont {Strohmaier}},
  \bibinfo {author} {\bibfnamefont {K.}~\bibnamefont {Gunter}}, \bibinfo
  {author} {\bibfnamefont {H.}~\bibnamefont {Moritz}}, \ and\ \bibinfo {author}
  {\bibfnamefont {T.}~\bibnamefont {Esslinger}},\ }\href@noop {} {\bibfield
  {journal} {\bibinfo  {journal} {Nature},\ }\textbf {\bibinfo {volume}
  {455}},\ \bibinfo {pages} {204} (\bibinfo {year} {2008})}\BibitemShut
  {NoStop}%
\bibitem [{\citenamefont {Ott}\ \emph {et~al.}(2004)\citenamefont {Ott},
  \citenamefont {de~Mirandes}, \citenamefont {Ferlaino}, \citenamefont {Roati},
  \citenamefont {Modugno},\ and\ \citenamefont {Inguscio}}]{collision_tran}%
  \BibitemOpen
  \bibfield  {author} {\bibinfo {author} {\bibfnamefont {H.}~\bibnamefont
  {Ott}}, \bibinfo {author} {\bibfnamefont {E.}~\bibnamefont {de~Mirandes}},
  \bibinfo {author} {\bibfnamefont {F.}~\bibnamefont {Ferlaino}}, \bibinfo
  {author} {\bibfnamefont {G.}~\bibnamefont {Roati}}, \bibinfo {author}
  {\bibfnamefont {G.}~\bibnamefont {Modugno}}, \ and\ \bibinfo {author}
  {\bibfnamefont {M.}~\bibnamefont {Inguscio}},\ }\href@noop {} {\bibfield
  {journal} {\bibinfo  {journal} {Phys. Rev. Lett.},\ }\textbf {\bibinfo
  {volume} {92}},\ \bibinfo {pages} {160601} (\bibinfo {year}
  {2004})}\BibitemShut {NoStop}%
\bibitem [{\citenamefont {Strohmaier}\ \emph {et~al.}(2007)\citenamefont
  {Strohmaier}, \citenamefont {Takasu}, \citenamefont {Gunter}, \citenamefont
  {Jordens}, \citenamefont {Kohl}, \citenamefont {Moritz},\ and\ \citenamefont
  {Esslinger}}]{interaction_tran}%
  \BibitemOpen
  \bibfield  {author} {\bibinfo {author} {\bibfnamefont {N.}~\bibnamefont
  {Strohmaier}}, \bibinfo {author} {\bibfnamefont {Y.}~\bibnamefont {Takasu}},
  \bibinfo {author} {\bibfnamefont {K.}~\bibnamefont {Gunter}}, \bibinfo
  {author} {\bibfnamefont {R.}~\bibnamefont {Jordens}}, \bibinfo {author}
  {\bibfnamefont {M.}~\bibnamefont {Kohl}}, \bibinfo {author} {\bibfnamefont
  {H.}~\bibnamefont {Moritz}}, \ and\ \bibinfo {author} {\bibfnamefont
  {T.}~\bibnamefont {Esslinger}},\ }\href@noop {} {\bibfield  {journal}
  {\bibinfo  {journal} {Phys. Rev. Lett.},\ }\textbf {\bibinfo {volume} {99}},\
  \bibinfo {pages} {220601} (\bibinfo {year} {2007})}\BibitemShut {NoStop}%
\bibitem [{\citenamefont {Hung}\ \emph {et~al.}(2010)\citenamefont {Hung},
  \citenamefont {Zhang}, \citenamefont {Gemelke},\ and\ \citenamefont
  {Chin}}]{slowmasstran}%
  \BibitemOpen
  \bibfield  {author} {\bibinfo {author} {\bibfnamefont {C.~L.}\ \bibnamefont
  {Hung}}, \bibinfo {author} {\bibfnamefont {X.}~\bibnamefont {Zhang}},
  \bibinfo {author} {\bibfnamefont {N.}~\bibnamefont {Gemelke}}, \ and\
  \bibinfo {author} {\bibfnamefont {C.}~\bibnamefont {Chin}},\ }\href@noop {}
  {\bibfield  {journal} {\bibinfo  {journal} {Phys. Rev. Lett.},\ }\textbf
  {\bibinfo {volume} {104}},\ \bibinfo {pages} {160403} (\bibinfo {year}
  {2010})}\BibitemShut {NoStop}%
\bibitem [{\citenamefont {Simon}\ \emph {et~al.}(2011)\citenamefont {Simon},
  \citenamefont {Bakr}, \citenamefont {Ma}, \citenamefont {Tai}, \citenamefont
  {Preiss},\ and\ \citenamefont {Greiner}}]{Greiner_spin}%
  \BibitemOpen
  \bibfield  {author} {\bibinfo {author} {\bibfnamefont {J.}~\bibnamefont
  {Simon}}, \bibinfo {author} {\bibfnamefont {W.~S.}\ \bibnamefont {Bakr}},
  \bibinfo {author} {\bibfnamefont {R.}~\bibnamefont {Ma}}, \bibinfo {author}
  {\bibfnamefont {M.~E.}\ \bibnamefont {Tai}}, \bibinfo {author} {\bibfnamefont
  {P.~M.}\ \bibnamefont {Preiss}}, \ and\ \bibinfo {author} {\bibfnamefont
  {M.}~\bibnamefont {Greiner}},\ }\href@noop {} {\bibfield  {journal} {\bibinfo
   {journal} {Nature},\ }\textbf {\bibinfo {volume} {472}},\ \bibinfo {pages}
  {307} (\bibinfo {year} {2011})}\BibitemShut {NoStop}%
\bibitem [{\citenamefont {Pepino}\ \emph {et~al.}(2009)\citenamefont {Pepino},
  \citenamefont {Cooper}, \citenamefont {Anderson},\ and\ \citenamefont
  {Holland}}]{atomtronics}%
  \BibitemOpen
  \bibfield  {author} {\bibinfo {author} {\bibfnamefont {R.~A.}\ \bibnamefont
  {Pepino}}, \bibinfo {author} {\bibfnamefont {J.}~\bibnamefont {Cooper}},
  \bibinfo {author} {\bibfnamefont {D.~Z.}\ \bibnamefont {Anderson}}, \ and\
  \bibinfo {author} {\bibfnamefont {M.~J.}\ \bibnamefont {Holland}},\
  }\href@noop {} {\bibfield  {journal} {\bibinfo  {journal} {Phys. Rev.
  Lett.},\ }\textbf {\bibinfo {volume} {103}},\ \bibinfo {pages} {140405}
  (\bibinfo {year} {2009})}\BibitemShut {NoStop}%
\bibitem [{\citenamefont {Landauer}(1957)}]{Landauer}%
  \BibitemOpen
  \bibfield  {author} {\bibinfo {author} {\bibfnamefont {R.}~\bibnamefont
  {Landauer}},\ }\href@noop {} {\bibfield  {journal} {\bibinfo  {journal} {IBM
  J. Res. Dev.},\ }\textbf {\bibinfo {volume} {1}},\ \bibinfo {pages} {223}
  (\bibinfo {year} {1957})}\BibitemShut {NoStop}%
\bibitem [{\citenamefont {Di~Ventra}\ and\ \citenamefont
  {Todorov}(2004)}]{Micro}%
  \BibitemOpen
  \bibfield  {author} {\bibinfo {author} {\bibfnamefont {M.}~\bibnamefont
  {Di~Ventra}}\ and\ \bibinfo {author} {\bibfnamefont {T.}~\bibnamefont
  {Todorov}},\ }\href@noop {} {\bibfield  {journal} {\bibinfo  {journal} {J.
  Phys. Cond Matt.},\ }\textbf {\bibinfo {volume} {16}},\ \bibinfo {pages}
  {8025} (\bibinfo {year} {2004})}\BibitemShut {NoStop}%
\bibitem [{\citenamefont {Bushong}\ \emph {et~al.}(2005)\citenamefont
  {Bushong}, \citenamefont {Sai},\ and\ \citenamefont {Di~Ventra}}]{MCE}%
  \BibitemOpen
  \bibfield  {author} {\bibinfo {author} {\bibfnamefont {N.}~\bibnamefont
  {Bushong}}, \bibinfo {author} {\bibfnamefont {N.}~\bibnamefont {Sai}}, \ and\
  \bibinfo {author} {\bibfnamefont {M.}~\bibnamefont {Di~Ventra}},\ }\href@noop
  {} {\bibfield  {journal} {\bibinfo  {journal} {Nano Lett.},\ }\textbf
  {\bibinfo {volume} {5}},\ \bibinfo {pages} {2569} (\bibinfo {year}
  {2005})}\BibitemShut {NoStop}%
\bibitem [{\citenamefont {Di~Ventra}(2008)}]{Maxbook}%
  \BibitemOpen
  \bibfield  {author} {\bibinfo {author} {\bibfnamefont {M.}~\bibnamefont
  {Di~Ventra}},\ }\href@noop {} {\emph {\bibinfo {title} {Electrical Transport
  in Nanoscale Systems}}}\ (\bibinfo  {publisher} {Cambridge University
  Press},\ \bibinfo {year} {2008})\BibitemShut {NoStop}%
\bibitem [{mun()}]{munote}%
  \BibitemOpen
  \href@noop {} {}\bibinfo {note} {At finite $T$, $\mu$ is determined by
  $n=\sum_{k}n_{k}/N$ for the initial state. Moreover, while the QSSC still
  occurs at higher $T$ in our model, experiments would need to be done at
  moderately low $T$ in order to retain atoms in the lowest energy
  band.}\BibitemShut {Stop}%
\bibitem [{bos()}]{bosonTnote}%
  \BibitemOpen
  \href@noop {} {}\bibinfo {note} {As $T$ increases, the initial boson
  distribution spreads and a QSSC can survive at high $T$.}\BibitemShut {Stop}%
\bibitem [{\citenamefont {Ho}\ and\ \citenamefont {Zhou}(2007)}]{Holattice}%
  \BibitemOpen
  \bibfield  {author} {\bibinfo {author} {\bibfnamefont {T.~L.}\ \bibnamefont
  {Ho}}\ and\ \bibinfo {author} {\bibfnamefont {Q.}~\bibnamefont {Zhou}},\
  }\href@noop {} {\bibfield  {journal} {\bibinfo  {journal} {Phys. Rev.
  Lett.},\ }\textbf {\bibinfo {volume} {99}},\ \bibinfo {pages} {120404}
  (\bibinfo {year} {2007})}\BibitemShut {NoStop}%
\bibitem [{\citenamefont {Kajala}\ \emph {et~al.}(2011)\citenamefont {Kajala},
  \citenamefont {Massel},\ and\ \citenamefont {Torma}}]{TormaPRL11}%
  \BibitemOpen
  \bibfield  {author} {\bibinfo {author} {\bibfnamefont {J.}~\bibnamefont
  {Kajala}}, \bibinfo {author} {\bibfnamefont {F.}~\bibnamefont {Massel}}, \
  and\ \bibinfo {author} {\bibfnamefont {P.}~\bibnamefont {Torma}},\
  }\href@noop {} {\bibfield  {journal} {\bibinfo  {journal} {Phys. Rev.
  Lett.},\ }\textbf {\bibinfo {volume} {106}},\ \bibinfo {pages} {206401}
  (\bibinfo {year} {2011})}\BibitemShut {NoStop}%
\bibitem [{\citenamefont {Polkovnikov}\ \emph {et~al.}(2011)\citenamefont
  {Polkovnikov}, \citenamefont {Sengupta}, \citenamefont {Silva},\ and\
  \citenamefont {Vengalattore}}]{PolkovnikovRMP}%
  \BibitemOpen
  \bibfield  {author} {\bibinfo {author} {\bibfnamefont {A.}~\bibnamefont
  {Polkovnikov}}, \bibinfo {author} {\bibfnamefont {K.}~\bibnamefont
  {Sengupta}}, \bibinfo {author} {\bibfnamefont {A.}~\bibnamefont {Silva}}, \
  and\ \bibinfo {author} {\bibfnamefont {M.}~\bibnamefont {Vengalattore}},\
  }\href@noop {} {\bibfield  {journal} {\bibinfo  {journal} {Rev. Mod. Phys.},\
  }\textbf {\bibinfo {volume} {83}},\ \bibinfo {pages} {863} (\bibinfo {year}
  {2011})}\BibitemShut {NoStop}%
\end{thebibliography}
%

\end{document}